\documentclass[prb,twocolumn,showpacs]{revtex4}
\usepackage{graphicx}

\begin{document}

\title{$d$-wave superconductivity from electron-phonon interactions}

\author{J.P.Hague}
\affiliation{Dept. of Physics and Astronomy, University of Leicester, Leicester, LE1 7RH}
\affiliation{Dept. of Physics, Loughborough University, Loughborough, LE11 3TU}

\begin{abstract}
I examine electron-phonon mediated superconductivity in the
intermediate coupling and phonon frequency regime of the quasi-2D
Holstein model. I use an extended Migdal--Eliashberg theory which
includes vertex corrections and spatial fluctuations. I find a
$d$-wave superconducting state that is unique close to
half-filling. The order parameter undergoes a transition to $s$-wave
superconductivity on increasing filling. I explain how the inclusion
of both vertex corrections and spatial fluctuations is essential for
the prediction of a $d$-wave order parameter. I then discuss the
effects of a large Coulomb pseudopotential on the superconductivity
(such as is found in contemporary superconducting materials like the
cuprates), which results in the destruction of the $s$-wave states,
while leaving the $d$-wave states unmodified. {\bf Published as: Phys. Rev. B 73, 060503(R) (2006)}
\end{abstract}

\date{4th May 2005}

\maketitle

The discovery of high transition temperatures and a $d$-wave order
parameter in the cuprate superconductors are remarkable results and
have serious implications for the theory of superconductivity. The
presence of large Coulomb interactions in the cuprates which have the
potential to destroy conventional $s$-wave BCS states has prompted the
search for new mechanisms that can give rise to
superconductivity. However, electron-phonon mediated superconductivity
is still not well understood, especialy in lower dimensional
systems. In particular, the electron-phonon problem is particularly
difficult at intermediate couplings with large phonon frequency (such
as found in the cuprates) and the electron-phonon mechanism cannot be
fully ruled out. It is therefore of paramount importance to develop
new theories to understand electron-phonon mediated superconductivity
away from the BCS limit.

The assumption that electron-phonon interactions cannot lead to high
transition temperatures and unusual order parameters was made on the
basis of calculations from BCS theory, which is a very-weak-coupling
mean-field theory (although of course highly successful for pre-1980s
superconductors) \cite{bcs}. In the presence of strong Coulomb
interaction, the BCS $s$-wave transition temperature is vastly
reduced. However, the recent measurement of large couplings between
electrons and the lattice in the cuprate superconductors means that
extensions to the conventional theories of superconductivity are
required \cite{zhao1997a,lanzara2001a,mcqueeny1999a}. In particular,
low dimensionality, intermediate dimensionless coupling constants of
$\sim 1$ and large and active phonon frequencies of $\sim$ 75meV mean
that BCS or the more advanced Migdal--Eliashberg (ME) theory cannot be
applied. In fact, the large coupling constant and a propensity for
strong renormalization in 2D systems, indicate that the bare
unrenormalized phonon frequency could be several times greater than
the measured 75 meV \cite{hague2003a}.

Here I apply the dynamical cluster approximation (DCA) to introduce a
fully self-consistent momentum-dependent self-energy to the
electron-phonon problem
\cite{hettler1998a,maier2004,hague2003a,hague2005d}. Short ranged
spatial fluctuations and lowest order vertex corrections are included,
allowing the sequence of phonon absorption and emission to be
reordered once. In particular, the theory used here is second order in
the effective electron-electron coupling $U=-g^2/M\omega_0^2$, which
provides the correct weak coupling limit from small to large phonon
frequencies \footnote{I also note the extensions to Eliashberg theory
carried out by Grimaldi \emph{et al.} \cite{grimaldi1995}.}. In this
paper, I include symmetry broken states in the anomalous self energy
to investigate unconventional order parameters such as $d$-wave. No
assumptions are made in advance about the form of the order parameter.

DCA \cite{hettler1998a,hettler2000a,maier2004} is an extension to the
dynamical mean-field theory for the study of low dimensional
systems. To apply the DCA, the Brillouin zone is divided into $N_C$
subzones within which the self-energy is assumed to be momentum
independent, and cluster Green functions are determined by averaging
over the momentum states in each subzone. This leads to spatial
fluctuations with characteristic range, $N_{c}^{1/D}$. In this paper,
$N_{c}=4$ is used throughout. This puts an upper bound on the strength
of the superconductivity, which is expected to be reduced in larger
cluster sizes \cite{jarrell2001a}. To examine superconducting states,
DCA is extended within the Nambu formalism \cite{maier2004,
hague2005d}. Green functions and self-energies are described by
$2\times 2$ matrices, with off diagonal terms relating to the
superconducting states. The self-consistent condition is:
\begin{equation}
G(\mathbf{K},i\omega_n)=\int_{-\infty}^{\infty}d\epsilon\frac{{\mathcal{D}}_i(\epsilon)(\zeta(\mathbf{K}_i,i\omega_n)-\epsilon)}{|\zeta(\mathbf{K}_i,i\omega_n)-\epsilon|^2+\phi(\mathbf{K}_i,i\omega_n)^{2}}
\label{eqn:grnsfnsc}
\end{equation}
\begin{equation}
F(\mathbf{K},i\omega_n)=-\int_{-\infty}^{\infty}d\epsilon\frac{{\mathcal{D}}_i(\epsilon)\phi(\mathbf{K}_i,i\omega_n)}{|\zeta(\mathbf{K}_i,i\omega_n)-\epsilon|^2+\phi(\mathbf{K}_i,i\omega_n)^{2}}
\label{eqn:grnsfnscanom}
\end{equation}
where
$\zeta(\mathbf{K}_i,i\omega_n)=i\omega_n+\mu-\Sigma(\mathbf{K}_i,i\omega_n)$,
$\mu$ is the chemical potential, $\omega_n$ are the Fermionic
Matsubara frequencies, $\phi(\mathbf{K},i\omega)$ is the anomalous
self energy and $\Sigma(\mathbf{K},i\omega)$ is the normal self
energy. $G(\mathbf{K},i\omega_n)$ must obey the lattice symmetry. In
contrast, it is only $|F(\mathbf{K},i\omega_n)|$ which is constrained
by this condition, since $\phi$ is squared in the denominator of
Eqn. \ref{eqn:grnsfnsc}. Therefore the sign of $\phi$ can
change. For instance, if the anomalous self energy has the rotational
symmetry $\phi(\pi,0)=-\phi(0,\pi)$, the on-diagonal Green function,
which represents the electron propagation retains the correct lattice
symmetry $G(\pi,0)=G(0,\pi)$. Therefore, only inversion symmetry is
required of the anomalous Green function representing
superconducting pairs and the anomalous self energy.

\begin{figure}[t]
\includegraphics[width=60mm]{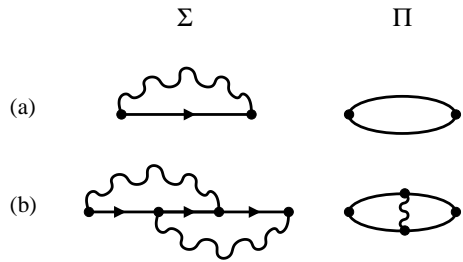}
\caption{Diagrammatic representation of the current
  approximation. Series (a) represents the vertex-neglected theory
  which corresponds to the Migdal--Eliashberg approach, valid when the
  phonon energy $\omega_0$ and electron-phonon coupling $U$ are small
  compared to the Fermi energy. Series (b) represents additional
  diagrams for the vertex corrected theory. The phonon self energies
  are labeled with $\Pi$, and $\Sigma$ denotes the electron
  self-energies. Lines represent the full electron Green function
  and wavy lines the full phonon Green function.}
\label{fig:feynmandiag}
\end{figure}

Here I examine the Holstein model \cite{holstein1959} of electron-phonon interactions. It treats
phonons as nuclei vibrating in a time-averaged harmonic potential
(representing the interactions between all nuclei), i.e. only one
frequency $\omega_0$ is considered. The phonons couple to the local
electron density via a momentum-independent coupling constant $g$
\cite{holstein1959}.
\begin{eqnarray}
H & =-\sum_{<ij>\sigma}t c^{\dagger}_{i\sigma}c_{j\sigma}+\sum_{i\sigma} n_{i\sigma} (gr_i-\mu) \nonumber\\
& +\sum_i\left( \frac{M\omega_{0}^2r_i^2}{2}+\frac{p_i^2}{2M}\right)
\end{eqnarray}
The first term in this Hamiltonian represents hopping of electrons
between neighboring sites and has a dispersion
$\epsilon_{\mathbf{k}}=-2t\sum_{i=1}^{D}\cos(k_{i})$. The second term couples
the local ion displacement, \( r_{i} \) to the local electron
density. The last term is the bare phonon Hamiltonian, i.e. a simple
harmonic oscillator. The creation and annihilation of electrons is
represented by \( c^{\dagger }_{i} \)(\( c_{i} \)), \( p_{i} \) is the
ion momentum and \( M \) the ion mass. The effective electron-electron
interaction is,
\begin{equation}
\label{eqn:phononfn}
U(i\omega _{s})=\frac{U\omega_{0}^{2}}{\omega _{s}^{2}+\omega _{0}^{2}}
\end{equation}
where, $\omega _{s}=2\pi sT$, $s$ is an integer and
$U=-g^2/M\omega_0^2$ represents the magnitude of the effective
electron-electron coupling. $D=2$ with $t=0.25$, resulting in a
non-interacting band width $W=2$. A small interplanar hopping
$t_{\perp}=0.01$ is included. This is necessary to stabilise
superconductivity, which is not permitted in a pure 2D system
\cite{hohenberg}.

Perturbation theory in the effective electron-electron interaction
(Fig. \ref{fig:feynmandiag}) is applied to second order in $U$,
using a skeleton expansion. The electron self-energy has two terms,
$\Sigma_{\mathrm{ME}}(\omega,\mathbf{K})$ neglects vertex corrections
(Fig. \ref{fig:feynmandiag}(a)), and
$\Sigma_{\mathrm{VC}}(\omega,\mathbf{K})$ corresponds to the vertex
corrected case (Fig.
\ref{fig:feynmandiag}(b)). $\Pi_{\mathrm{ME}}(\omega,\mathbf{K})$ and
$\Pi_{\mathrm{VC}}(\omega,\mathbf{K})$ correspond to the equivalent
phonon self energies. At large phonon frequencies, all second order
diagrams including $\Sigma_{VC}$ are essential for the correct
description of the weak coupling limit.

The phonon propagator $D(z,\mathbf{K})$ is calculated from,
\begin{equation}
D(i\omega_s,\mathbf{K})=\frac{\omega_0^2}{\omega_s^2+\omega_0^2-\Pi(i\omega_s,\mathbf{K})}
\label{eqn:phonprop}
\end{equation}
and the Green function from equations \ref{eqn:grnsfnsc} and
\ref{eqn:grnsfnscanom}.
$\underline{\Sigma}=\underline{\Sigma}_{\mathrm{ME}}+\underline{\Sigma}_{\mathrm{VC}}$
and $\Pi=\Pi_{\mathrm{ME}}+\Pi_{\mathrm{VC}}$.  Details of the
translation of the diagrams in Fig. \ref{fig:feynmandiag} and the
iteration procedure can be found in Ref.
\onlinecite{hague2005d}. Calculations are carried out along the
Matsubara axis, with sufficient Matsubara points for an accurate
calculation. The equations were iterated until the normal and
anomalous self-energies converged to an accuracy of approximately 1
part in $10^3$.

Since the anomalous Green function is proportional to the anomalous
self energy, initializing the problem with the non-interacting Green
function leads to a non-superconducting (normal) state. A constant
superconducting field with $d$-wave symmetry was applied to the system
to induce superconductivity. \emph{The
external field was then completely removed. Iteration continued
without the field until convergence}. This solution was then used
to initialize self-consistency for other similar values of the
parameters. The symmetry conditions used in Refs
\onlinecite{hague2003a} and \onlinecite{hague2005d} have been relaxed
to reflect the additional breaking of the anomalous lattice symmetry
in the $d$-wave state. This does not affect the normal state Green
function, but does affect the anomalous state Green function.

\begin{figure}
\includegraphics[width=60mm,angle=270]{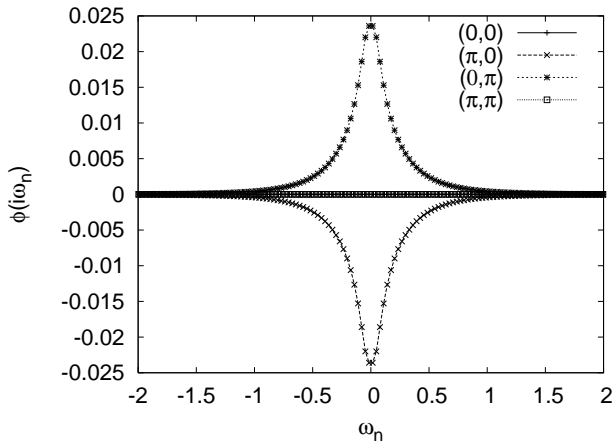}
\caption{Anomalous self-energy at half-filling. The anomalous self
energy is real. It is clear that $\phi(\pi,0)=-\phi(0,\pi)$. This is
characteristic of $d$-wave order. Similarly, the electron self energy
has the correct lattice symmetry $\Sigma(\pi,0)=\Sigma(0,\pi)$, which
was not imposed from the outset. The gap function is related to the anomalous self energy via $\phi(i\omega_n)/Z(i\omega_n)$.}
\label{fig:selfenergy}
\end{figure}

In Fig. \ref{fig:selfenergy}, the anomalous self energy is examined
for $n=1.0$ (half-filling). The striking feature is that stable $d$-wave
superconductivity is found. This is manifested through a change in
sign of the anomalous self energy, which is negative at the $(\pi,0)$
point and positive at the $(0,\pi)$ point. The electron Green
function (equation \ref{eqn:grnsfnsc}) depends on $\phi^2$, so
causality and lattice symmetry are maintained. Since the gap function
$\phi(i\omega_n)/Z(i\omega_n)$ is directly proportional to
$\phi(i\omega_n)$, and
$Z(i\omega_n,\mathbf{K}_{(\pi,0)})=Z(i\omega_n,\mathbf{K}_{(0,\pi)})$,
then the sign of the order parameter i.e. the sign of the
superconducting gap changes under $90^{o}$
rotation. $Z(i\omega_n)=1-\Sigma(i\omega_n)/i\omega_n$.

\begin{figure}
\includegraphics[width=55mm,angle=270]{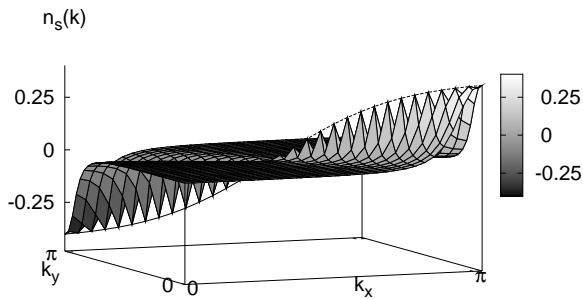}
\caption{Variation of superconducting (anomalous) pairing density
across the Brillouin zone. $n_s(\mathbf{k})=T\sum_n
F(i\omega_n,\mathbf{k})$. $U=0.6, \omega_0=0.4, n=1$ and
$T=0.005$. The $d$-wave order can be seen very clearly, with a change
in sign on 90$^o$ rotation and a node situated at the $(\pi/2,\pi/2)$
point. The largest anomalous (superconducting) densities are at the
$(\pi,0)$ and $(0,\pi)$ points.}
\label{fig:effectcluster}
\end{figure}

\begin{figure}
\includegraphics[width=60mm,angle=270]{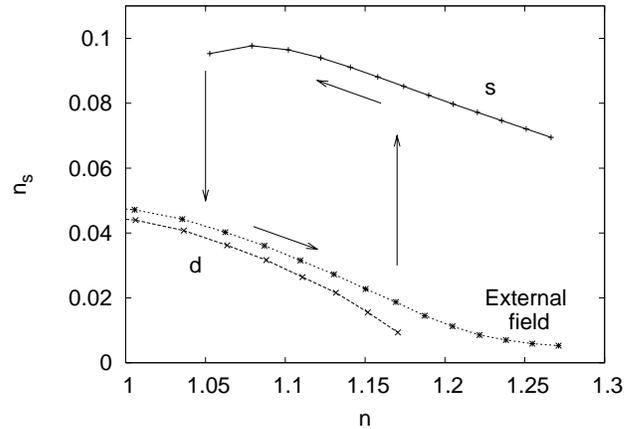}
\caption{Hysteresis of the superconducting order
  parameters. $n_s=\sum_{\mathbf{K}}|n_s(\mathbf{K}_i)|$. Starting
  from a $d$-wave state at half-filling, increasing the chemical
  potential increases the filling and decreases the $d$-wave
  order. Eventually, at $n=1.18$ the system changes to an $s$-wave
  state. On return from large filling, the $s$-wave superconductivity
  is persistent to a low filling of $n=1.04$, before spontaneously
  reverting to a $d$-wave state. The system is highly susceptible to
  $d$-wave order, and application of a very small external
  superconducting field to an $s$-wave state results in a $d$-wave
  state. Note that $d$- and $s$-wave channels are coupled in the
  higher order theory, so the transition can take place spontaneously,
  unlike in the standard gap equations.}
\label{fig:hysteresis}
\end{figure}

Figure \ref{fig:effectcluster} shows the variation of superconducting
pairing across the Brillouin zone. $n_s(\mathbf{k})=T\sum_n
F(i\omega_n,\mathbf{k})$. $U=0.6, \omega_0=0.4, n=1$ and
$T=0.005$. The $d$-wave order can be seen very clearly. The largest
anomalous densities are at the $(\pi,0)$ and $(0,\pi)$ points, with a
node situated at the $(\pi/2,\pi/2)$ point and a sign change on
90$^{o}$ rotation. Pairing clearly occurs between electrons close to the Fermi surface.

So far, the model has been analyzed at half filling. Figure
\ref{fig:hysteresis} demonstrates the evolution of the order parameter
as the number of holes is first increased, and then decreased. The
total magnitude of the anomalous density,
$n_s=\sum_{\mathbf{K}}|n_s(\mathbf{K}_i)|$ is examined. When the
number of holes is increased, stable $d$-wave order persists to a
filling of $n=1.18$, while decreasing monotonically. At the critical
point, there is a spontaneous transition to $s$-wave order. Starting
from a high filling, and reducing the number of holes, there is a
spontaneous transition from $s$ to $d$-wave order at $n=1.04$. There
is therefore hysteresis associated with the self-consistent
solution. It is reassuring that the $d$-wave state can be induced
without the need for the external field. As previously established,
$s$-wave order does not exist at half-filling as a mainfestation of
Hohenberg's theorem \cite{hague2005d}, so the computed $d$-wave order
at half-filling is the ground state of the model. It is interesting
that the $d$- and $s$-channels are able to coexist, considering that
the BCS channels are separate on a square lattice. This is due to the
vertex corrections, since the self consistent equations are no longer
linear in the gap function (the 1st order gap equation vanishes in the
$d$-wave case, leaving 2nd order terms as the leading contribution).

I finish with a brief discussion of Coulomb effects. In the Eliashberg
equations, a Coulomb pseudopotential may be added to the theory as,
\begin{equation}
\phi_C=U_C T\sum_{\mathbf{K},n}F(i\omega_n,\mathbf{K})
\end{equation}
It is easy to see the effect of $d$-wave order on this term. Since the
sign of the anomalous Green function is modulated, the average
effect of $d$-wave order is to nullify the Coulomb contribution to the
anomalous self-energy (i.e. $\phi_{Cd}=0$). This demonstrates that the
$d$-wave state is stable to Coulomb perturbations, presumably because
the pairs are distance separated. In contrast, the $s$-wave state is
not stable to Coulomb interaction, with a corresponding reduction of
the transition temperature ($T_C=0$ for $\lambda<\mu_C$). Thus,
such a Coulomb filter selects the $d$-wave state (see
e.g. Ref. \onlinecite{annett}). Since large local Coulomb repulsions are
present in the cuprates (and indeed most transition metal oxides),
then this mechanism seems the most likely to remove the
hysteresis. Without the Coulomb interactions, it is expected that the
$s$-wave state will dominate for $n>1.04$, since the anomalous order
is larger.

I note that a further consequence of strong Coulomb repulsion is antiferromagnetism
close to half-filling. Typically magnetic fluctuations act to suppress phonon mediated
superconducting order. As such, one might expect a suppression of
superconducting order close to half-filling, with a maximum away from
half filling. The current theory could be extended to include
additional anomalous Green functions related to antiferromagnetic
order. This would lead to a 4x4 Green function matrix.  A full
analysis of antiferromagnetism and the free energy will be carried out
at a later date.

\paragraph{Summary}

In this paper I have carried out simulations of the 2D Holstein model
in the superconducting state. Vertex corrections and spatial
fluctuations were included in the approximation for the
self-energy. The anomalous self energy and superconducting order
parameter were calculated. Remarkably, stable superconducting states with $d$-wave order were
found at half-filling. $d$-wave states persist
to $n=1.18$, where the symmetry of the parameter changes to
$s$-wave. Starting in the $s$-wave phase and reducing the filling,
$d$-wave states spontaneously appear at $n=1.04$.  The spontaneous
appearance of $d$-wave states in a model of electron-phonon
interactions is of particular interest, since it may negate the need
for novel pairing mechanisms in the cuprates \footnote{ On the basis of a screened electron-phonon interaction,
Abrikosov claims to have found stable $d$-wave states in a BCS like
theory \cite{abrikosova,abrikosovb}. However with an unscreened
Holstein potential, the transition temperature it the $d$-wave channel
given by the standard theory is zero. Also, the assumed order
parameter in his work does not clearly have $d$-wave
symmetry.}.

The inclusion of vertex corrections and spatial fluctuations was
essential to the emergence of the $d$-wave states in the Holstein
model, which indicates why BCS and ME calculations do not predict this
phenomenon. For very weak coupling, the off diagonal Eliashberg
self-energy has the form
$-UT\sum_{\mathbf{Q},n}F(i\omega_n,\mathbf{Q})D_0(i\omega_s-i\omega_n)$,
so it is clear (for the same reasons as the Coulomb pseudopotential)
that this diagram has no contribution in the $d$-wave
phase (the weak coupling phonon propagator is momentum independent for the Holstein model). Therefore, vertex corrections are the leading term in the weak
coupling limit. Furthermore, I have discussed the inclusion of Coulomb
states to lowest order, which act to destabilize the $s$-wave states,
while leaving the $d$-wave states unchanged. Since the Coulomb
pseudopotential has no effect then it is possible that electron-phonon
interactions are the mechanism inducing $d$-wave states in real
materials such as the cuprates. The Coulomb filtering mechanism works
for $p$-wave symmetry and higher, so it is possible that
electron-phonon interactions could explain many novel
superconductors. Certainly, such a mechanism cannot be ruled out. The doping dependence of the order qualitatively matches that of La$_{2-x}$Sr$_{x}$CuO$_{4}$ (here order extends to $x=0.18$, in the Cuprate to $x=0.3$). Antiferromagnetism is only present in the cuprate very close to half filling (up to approx $x=0.02$), and on a mean-field level does not interfere with the $d$-wave superconductivity at larger dopings.

It has been determined experimentally that strong electron-phonon
interactions and high phonon frequencies are clearly visible in the
electron and phonon band structures of the cuprates, and are therefore
an essential part of the physics
\cite{lanzara2001a,mcqueeny1999a}. Similar effects to those observed
in the cuprates are seen in the electron and phonon band structures of
the 2D Holstein model in the normal phase \cite{hague2003a}. It is
clearly of interest to determine whether other features and effects in
the cuprate superconductors could be explained with electron-phonon
interactions alone.

\paragraph{Acknowledgments}

I thank the University of Leicester for hospitality while carrying out this work. I thank E.M.L.Chung for useful
discussions. I am currently supported under EPSRC grant no. EP/C518365/1.

\end{document}